\documentclass[aps,prl,twocolumn,groupedaddress,showpacs]{revtex4}

\begin{document}

\title{Quantum nonlocality without hidden variables: An algorithmic approach}

\author{Hrvoje Nikoli\'c}
\affiliation{Theoretical Physics Division, Rudjer Bo\v{s}kovi\'{c}
Institute,
P.O.B. 180, HR-10002 Zagreb, Croatia.}
\email{hrvoje@thphys.irb.hr}

\date{\today}

\begin{abstract}
Is quantum mechanics (QM) local or nonlocal?
Different formulations/interpretations (FI) of QM, with or without hidden
variables, suggest different answers. Different FI's can be viewed as different 
algorithms, which leads us to propose an algorithmic definition of locality 
according to which a theory is local if and only if there exists at least one 
FI in which all irreducible elements of that FI are local. The fact that 
no such FI of QM is known strongly supports quantum nonlocality.
\end{abstract}


\maketitle

Is quantum mechanics (QM) local or nonlocal?
The no-hidden-variable theorems, such as those of Bell \cite{bell},
Greenberger, Horne, and Zeilinger \cite{GHZ}, and Hardy \cite{hardy}, 
prove that no local hidden variable theory can reproduce the predictions
of QM. This means that if hidden variables
(i.e., objective elements of reality existing even without measurements) 
exist, then they cannot be local. Some take this as an indication that
hidden variables cannot exist, while others take this as an indication
that QM is nonlocal. Thus, to definitely answer the question whether QM
is local or not, we need a definition of locality that does not rest
on the notion of (hypothetical) hidden variables.

One such widely used definition of locality is {\em signal} locality.
According to this definition, QM is local if it does not allow 
to send a signal (information) that propagates faster than light.
Clearly, according to this definition, QM is local. 
Nevertheless, although it is widely appreciated that QM obeys signal
locality, there is still no consensus regarding the question 
whether QM is local or not. In particular, nonlocal hidden-variable 
interpretations of QM, such as the Bohmian interpretation \cite{bohm},
also obey signal locality, but this does not stop the opponents of 
hidden-variable interpretations to
criticize such interpretations for not being local.
Thus, signal locality is not a completely satisfying definition of 
locality either.
To achieve a consensus regarding the question of quantum locality,
we obviously need a different, more general, definition 
of locality itself. 

Since the origin of all quantum nonlocalities can be reduced to
the entanglement between different particles, 
it is sometimes argued that there is nothing really nonlocal about
QM because particles need to perform a local interaction in order
to come into an entangled state. However, even if it is true
in practical experiments, it is not true in principle. Even 
the free Schr\"odinger equation of two noninteracting particles contains
entangled solutions, while no theoretical principle
of QM forbids such solutions. 

Last but not least, adherents of different interpretations of QM 
typically have different opinions on quantum locality.
For example, adherents of hidden-variable interpretations, 
conceiving that elements of objective reality must exist 
even if we do not measure them, typically find QM intrinsically
nonlocal. On the other hand, adherents of some interpretations
that explicitly deny the existence of a unique objective reality,
such as the many-world interpretation \cite{mw1,mw2,mw3} 
and the relational interpretation \cite{rov1,rov2}, argue that these 
interpretations save locality of QM. Still, by definition,
all these interpretations have the same observable predictions; 
they are all {\em observationally equivalent}. Thus,  
to achieve a consensus regarding the question of quantum locality,
it is very important to have a definition that does not depend
on the interpretation. 
 
Now we have rather strong constraints on a satisfying definition
of locality. On the one hand, it should not be a purely 
observational definition (e.g., like signal locality), 
because the adherents of hidden-variable 
interpretations will complain that observational locality
does not exclude the existence of nonlocal hidden variables.
On the other hand, it should be an interpretation-independent
definition, so that, in particular, it does not depend on 
whether the existence of hidden variables is assumed or not. 
How to fulfill both of these two somewhat opposite constraints? 
As the formalism of QM is the aspect of QM that is usually not regarded
as controversial, our idea is to 
use a {\em formal-theoretical} definition of locality.
 
For simplicity, we study two-particle systems with particle positions
${\bf x}$ and ${\bf y}$. Formal-theoretical elements
of the form $f({\bf x})$ or $f({\bf y})$ (where $f$ may be a 
function, an operator, or whatever) that depend only on
${\bf x}$ or only on ${\bf y}$ are said to be local.
By contrast, a formal-theoretical element of the form
$f({\bf x},{\bf y})$ depending on both ${\bf x}$ and ${\bf y}$
is said to be nonlocal. Clearly, the existence of 
nonlocal elements is a sign of possible nonlocality of the theory.
Indeed, the source of all nonlocal properties of QM can be 
traced back to the formal nonlocality of wave functions 
$\psi({\bf x},{\bf y})$, living in the configuration space, 
rather than in the ordinary 3-space. Nevertheless, the existence of 
nonlocal elements in a theory is certainly not yet the proof
of nonlocality of the theory. To see this, it is instructive 
to discuss some simple examples.  

Consider a system of two {\em classical} particles.
It is certainly possible to build a classical statistical ensemble
in which the positions of the particles are correlated, 
such that the probability distribution $P({\bf x},{\bf y})$
depends on both ${\bf x}$ and ${\bf y}$. Clearly, 
$P({\bf x},{\bf y})$ is a nonlocal element of the classical theory.
But does it mean that classical mechanics is nonlocal?
Certainly not. To see that this nonlocal element
does not necessarily imply nonlocality of the theory,
one may recall that classical mechanics can be formulated
in terms of hidden variables, which are particle positions 
existing even when they are not measured. (Note, however, 
that all predictions of classical mechanics can be
reproduced even without such hidden variables \cite{niknodet}.) 
For example, if particles have time-dependent trajectories
${\bf x}={\bf v}t$, ${\bf y}={\bf b}+{\bf v}t$, then
$P({\bf x},{\bf y})=\delta^3({\bf x}-{\bf y}+{\bf b})$.
Thus, although $P({\bf x},{\bf y})$ is nonlocal, this
nonlocal correlation between ${\bf x}$ and ${\bf y}$ can 
be recovered from hidden variables ${\bf x}$ and ${\bf y}$
that represent local elements of the theory.

The example above is certainly not sufficiently general
to prove locality of classical mechanics. For a general proof,
one needs a {\em general formulation} of two-particle classical
mechanics. One such formulation sufficiently general
for our purposes is the 
Hamilton-Jacobi formulation, where the central formal-theoretical
element is the nonlocal function $S({\bf x},{\bf y},t)$ satisfying
the Hamilton-Jacobi equation
\begin{equation}
\frac{(\nabla_x S)^2}{2m_1} + \frac{(\nabla_y S)^2}{2m_2}
+V_1({\bf x}) +V_2({\bf y})=-\partial_t S .
\end{equation}
Clearly, the Hamilton-Jacobi formulation is not local.
However, when the Hamilton-Jacobi equation is combined 
with another fundamental element of the Hamilton-Jacobi formulation
\begin{equation}\label{eomHJ}
m_1\dot{{\bf x}}(t)=\nabla_x S , \;\;\;
m_2\dot{{\bf y}}(t)=\nabla_y S ,
\end{equation}
then various different but observationally equivalent formulations of 
classical mechanics
can be obtained. Examples are the Hamiltonian and the Lagrangian
formulations, which are also nonlocal, because the fundamental
elements (Hamiltonians and Lagrangians) of these formulations 
are nonlocal.
Nevertheless, one of the formulations, the Newton formulation
consisting of the two equations 
\begin{equation}\label{eomcl}
m_1\ddot{{\bf x}}(t)=-\nabla_x V_1({\bf x}) , \;\;\;
m_2\ddot{{\bf y}}(t)=-\nabla_y V_2({\bf y}) ,
\end{equation}
does not contain any nonlocal elements. Thus, {\em the fact that there exists
at least one formulation of classical mechanics that does not
contain nonlocal elements proves that classical mechanics
is local!}

We also emphasize that the requirement of locality does not exclude
interactions between particles. The locality of interactions
is the most explicitly expressed in field theories, where 
a finite number of derivatives of a field $\phi({\bf x})$ 
is regarded as a local object.

Now let us turn back to QM.
The Hamilton-Jacobi formulation of classical mechanics above
is very similar to the Schr\"odinger formulation of 
QM. The Schr\"odinger formulation describing
the nonlocal wave function 
$\psi({\bf x},{\bf y},t)=R({\bf x},{\bf y},t) 
\exp[ iS({\bf x},{\bf y},t)/\hbar ]$ is indeed nonlocal.
By analogy with classical mechanics, one might hope 
that one could combine the Schr\"odinger equation with    
Eq.~(\ref{eomHJ}) to obtain an observationally equivalent formulation
that, analogously to the Newton formulation of classical mechanics, 
does not contain nonlocal elements. Indeed,  
the resulting formulation really turns out to be observationally 
equivalent to the Schr\"odinger formulation. Nevertheless, it 
does not eliminate all nonlocal elements from the theory.
Instead of the local Newton equations (\ref{eomcl}) one obtains 
nonlocal quantum analogues \cite{bohm}
\begin{eqnarray}\label{eomq}
m_1\ddot{{\bf x}}(t)=-\nabla_x [V_1({\bf x}) +Q({\bf x},{\bf y},t)] , 
\nonumber \\
m_2\ddot{{\bf y}}(t)=-\nabla_y [V_2({\bf y}) +Q({\bf x},{\bf y},t)] ,
\end{eqnarray}
where $Q({\bf x},{\bf y},t)$ is the nonlocal quantum potential \cite{bohm}.

The formulation of QM containing Eqs.~(\ref{eomq}) is known as the 
Bohmian interpretation \cite{bohm}. (We believe that, in general,
it is not possible to clearly distinguish
between the notions of ``formulations"
and ``interpretations", as long as both different formulations
and different interpretations are observationally equivalent.
For example, the Bohmian interpretation can be viewed as a 
practically useful formulation \cite{lopr}, while 
the Schr\"odinger formulation may be viewed as an interpretation
by someone who thinks of a wave function as a physical object
evolving with time. This is why we use the expressions 
``formulation" and ``interpretation" interchangeably.)
It is in fact the Bohmian formulation that inspired Bell 
to find his famous theorem on quantum nonlocality and hidden variables.
Nevertheless, this does not prove nonlocality of QM itself, as 
different formulations of QM without hidden variables are also possible.
 
Abstractly, a physical theory can be viewed as an equivalence class of 
{\em all}
observationally equivalent formulations/interpretations (FI) of that theory. 
If some property of a theory is not manifest in one FI, it may be manifest
in another. An FI of a theory is nothing but a set of 
irreducible elements (axioms, definitions, as well as primitive
``common-sense" objects that are not explicitly defined), 
which are mutually independent (no axiom or definition can be deduced
from other axioms and definitions) and
from which all other properties of the theory can be deduced. 
Even more abstractly, any such FI can be viewed as an {\em algorithm}
for determinination of reducible elements of the theory, 
such as specific observable predictions of physical theories. 
In the case of QM, these observable predictions are probabilities, 
such as the joint probability $P({\bf x},{\bf y})$ in a specific
physical configuration. The probabilities themselves as general
elements not attributed to a specific physical configuration
may or may not be irreducible elements of an FI. When they are not
irreducible elements in an FI,
we say that the probabilities are emergent in that FI. 
For example, the probabilities
are emergent in the deterministic formulation of classical mechanics,
in some versions of the Bohmian deterministic formulation of QM 
\cite{val,durr}, as well as in some versions of the 
many-world interpretation \cite{deut,wall,saun} of QM. 
However, in most FI's of QM the probabilities are
fundamental, i.e., correspond to some irreducible elements of the FI.

From the observations above we see that locality of an FI
is manifest if and only if all irreducible elements of that FI are local.   
Since the theory is the equivalence class of all observationally 
equivalent FI's of that theory, it motivates us to define  
locality of the theory as follows:
{\it A theory is local if and only if 
there exists an FI of the theory in which all irreducible elements are local.}
Since any FI can be viewed more abstractly as an algorithm, we refer
to this definition of locality as {\em algorithmic locality}.
Indeed, the definition above can be rephrased more concisely as follows:
{\it A theory is local if and only if it can be represented by a 
local algorithm.} 

The main motivation for such an abstract definition of locality
is to provide a resolution of the old controversy regarding 
the question whether QM is local or nonlocal. 
As a tentative answer to this 
question, we propose the following {\em algorithmic nonlocality conjecture}:
{\it No local algorithm can reproduce the probabilistic predictions
of QM.}

Note that this conjecture sounds very similar to the already 
proven hidden-variable nonlocality theorem saying that 
{\em no local hidden variables can reproduce the probabilistic predictions
of QM.} Indeed, all abstract irreducible elements of an algorithm
that are not directly observable may be viewed as ``hidden variables".
However, the notion of hidden variables actually has a more specific 
meaning,
denoting elements that are {\em observable}, but defined 
even when they are not observed. Thus, our 
algorithmic nonlocality conjecture is much more general (and thus much
more difficult to prove) than the hidden-variable nonlocality theorem.

Although we do not (yet) have a proof of the algorithmic nonlocality
conjecture, there is a lot of evidence supporting it. In the following we
present some of this evidence by supporting it by some specific FI's of QM. 

It is often argued that QM is essentially only about (local) information 
(see, e.g., \cite{zeil,peres} and references therein).
The relational interpretation \cite{rov1,rov2} is also a variant of this
idea.
If QM is only about local information, then signal locality
(referring to the propagation of information) is a good criterion of
locality. 
However, if QM is really only about information, then it is
reasonable to expect that there should exist a formulation of QM 
in which {\em all} irreducible elements refer to information itself.
Such a formulation should {\em not} contain auxiliary elements 
(such as wave functions or Hamiltonians) from which information can be 
deduced, but which do not represent information itself. Unfortunately,
no such formulation is known and it does not seem that such a
formulation is possible. From this we conclude that, in the algorithmic 
sense, QM does {\em not} seem to be only about information. Consequently, 
signal locality does not support algorithmic locality.

A variant of the information-theoretic interpretation of QM 
contains a claim that QM is essentially only about correlations
\cite{mermcor}. However, we know that quantum correlations 
may be nonlocal (typically, EPR correlations). Thus, even if it is 
possible to formulate QM only in terms of correlations (which we doubt), 
it is almost a tautology that these elements of formulation must be nonlocal, 
which supports algorithmic nonlocality.

One of the origins of quantum nonlocalities is the concept
of wave-function collapse. Some no-hidden-variable interpretations,
such as the consistent-histories interpretation 
\cite{conshist1,conshist2,conshist3}
and the many-world interpretation \cite{mw1,mw2,mw3}, completely 
eliminate wave-function collapse from the fundamental formulation
of the theory. Nevertheless, these interpretations do not
eliminate all nonlocal elements from the fundamental 
irreducible formulation.
In the consistent-histories interpretation, the probabilities 
of different histories are calculated with the aid of a 
nonlocal state described by a density matrix in the Heisenberg picture.
The many-world interpretation reformulates a quantum superposition
$\psi({\bf x},{\bf y})=\sum_{a} f_a({\bf x})h_a({\bf y})$
as a collection of different ``worlds" 
$\{ f_a({\bf x})h_a({\bf y}) \}$. Nevertheless, the single ``worlds"
$f_a({\bf x})h_a({\bf y})$ are still nonlocal objects.
The many-world interpretation  
cannot be further reformulated in terms of 
local ``worlds" $\{ f_a({\bf x}), h_a({\bf y}) \}$ only, because 
the specific pairings in which $f_1$ is paired with $h_1$, $f_2$ with $h_2$, 
and so on, are an important part of the formulation that allows 
the correct prediction of observable joint probabilities.  

A local formulation of QM should not contain nonlocal states
(represented, e.g., by nonlocal wave functions $\psi({\bf x},{\bf y})$ 
or nonlocal density matrices) as irreducible elements of the
formulation. The only such formulation we are aware of is the Nelson
interpretation \cite{nels,smol}, which, indeed, eliminates both local
and nonlocal wave functions from the fundamental formulation
of the theory. However, it is a hidden-variable interpretation, which, of
course, is nonlocal. 

From the discussion above, we can conclude that no present formulation of QM
obeys algorithmic locality, which strongly supports the 
algorithmic nonlocality conjecture. Nevertheless, let us assume that
this conjecture is wrong, that a local formulation, though yet unknown,
still exists. In particular, it would mean that all presently known FI's
of QM could be viewed as incomplete, which could also be thought of
as an algorithmic version of the old EPR argument \cite{EPR}.
An interesting question is: Would that mean that 
hidden-variable formulations of QM, being observationally
equivalent to this hypothetical local formulation, should be considered
local? According to our definition of algorithmic locality, 
the answer is -- yes. In this sense, hidden-variable 
interpretations of QM could be local. However, we emphasize that
this is true only if these hidden-variable formulations really {\em are}
observationally equivalent to the local formulation. 
In this regard, we note that some variants of the Bohmian interpretation
may, under certain circumstances, observationally differ from other
formulations. The examples of such circumstances are 
the early universe \cite{val2} 
and certain relativistic conditions \cite{nikfpl3}.   

As a final aside remark, not essential for the main thesis of this 
work, but interesting for a discussion of quantum
nonlocalities without hidden variables in a wider context, 
we note that some aspects
of string theory possess some surprising nonlocal features \cite{nlstr1,nlstr2}
that have no analogs in the quantum theory of particles or fields.
The origin of these nonlocalities turns out to lie in certain
{\em dualities} of string theory. Namely, in two different string theories
defined on different background manifolds,
some observable properties (e.g., the spectrum
of the Hamiltonian) turn out to be completely identical.
As recognized and illustrated on the example of T-duality in \cite{niktdual}, 
these nonlocalities 
arise precisely owing to the (tacit) assumption
that hidden variables do {\em not} exist. Namely, the nonlocalities arise
from the identification of theories that coincide in 
certain observable properties, but such an identification
makes sense only if all other unobserved properties are unphysical. 
On the other hand, the assumption of the existence of hidden variables 
means that unobserved properties may also be physical, which then removes 
dualities and corresponding nonlocalities at the fundamental
hidden-variable level. Nevertheless, this does not remove 
nonlocalities completely, as all hidden-variable formulations of QM
must be nonlocal. These theoretical observations are also indirectly 
related to the conjecture 
that quantum nonlocalities cannot be completely removed by reformulations
of the theory. 

To summarize, in this paper we have motivated a new definition of 
locality called algorithmic locality and presented evidence for the 
validity of the algorithmic nonlocality conjecture, according to
which no local algorithm can reproduce the probabilistic predictions
of QM. Essentially, this conjecture seems to be valid because any
formulation of QM seems to require nonlocal wave functions
(or some other nonlocal substitute for them) as one of the
fundamental irreducible elements of the formulation.

This work was supported by the Ministry of Science and Technology of the
Republic of Croatia.

\end{document}